\documentclass[preprint,showpacs,preprintnumbers,amsmath,amssymb,pra,superscriptaddress]{revtex4}

\usepackage[dvips]{graphicx} 
\usepackage{amsfonts}
\usepackage{amscd}
\usepackage{amsmath}    
\usepackage{enumerate}

\begin{document}

\title{Experimental quantum key distribution with an untrusted source}

\author{Xiang Peng}
\author{Hao Jiang}
\author{Bingjie Xu}
\affiliation{CREAM Group, State Key Laboratory of  Advanced  Optical Communication Systems and
Networks (Peking University) and Institute of Quantum Electronics, School of Electronics
Engineering and Computer Science, Peking University, Beijing 100871, PR China}

\author{Xiongfeng Ma}
\affiliation{Department of Physics and Astronomy, Institute of Quantum Computing, University of
Waterloo, 200 University Avenue West, Waterloo N2L 3G1, Ontario, Canada}
\author{Hong Guo}
\email{hongguo@pku.edu.cn} \affiliation{CREAM Group, State Key Laboratory of  Advanced  Optical
Communication Systems and Networks (Peking University) and Institute of Quantum Electronics,
School of Electronics Engineering and Computer Science, Peking University, Beijing 100871, PR
China}

\begin{abstract}
The photon statistics of a quantum key distribution (QKD) source is crucial for security
analysis. In this paper, we propose a practical method, with only a beam splitter and
photodetector involved, to monitor the photon statistics of a QKD source. By implementing in a
Plug$\&$Play QKD system, we show that the method is highly practical. The final secure key rate
is 52 bit/s, comparing to 78 bit/s when the source is treated as a trusted source.
\end{abstract}


\maketitle

Quantum key distribution (QKD) can establish a secret key between two parties, Alice and Bob,
by a quantum channel and an authenticated classical channel \cite{BB_84}. The unconditional
security of QKD has been proven even when imperfect devices are used
\cite{GLLP_04,RevQKD_Lut_08}.

A QKD system is composed of three parts: source, channel and detection. In the GLLP security
analysis \cite{GLLP_04}, the characteristics of these three parts are assumed to be fixed or
measured and known to Alice and Bob. To guarantee the security of a real QKD experiment, one
needs to carefully verify these assumptions.




The decoy state method \cite{Hwang_03,Decoy_05,Wang_05} is proposed to characterize the
properties of a QKD channel. Since a perfect single photon source is currently not available,
an imperfect single photon source, such as a weak coherent state source, is used in real QKD
setups. These imperfect single photon sources may contain some components that are not secure
for QKD use, e.g., multi-photon state. When Alice and Bob are not able to monitor the channel
properties, they have to pessimistically assume that all the losses and errors come from the
single photon components. In this case, the QKD performance is very limited. Fortunately, with
the decoy states, one can estimate the transmission efficiency of the single photon state
accurately and improve the QKD performance dramatically. Note that in the security proof of
decoy state QKD \cite{Decoy_05}, the photon statistics of the source is assumed to be fixed and
known to Alice and Bob. The main objective of the paper is to show how to monitor the photon
source.

For the detection part, the squash model is assumed in the GLLP security analysis.
With the squash model, one can assume that Eve, the eavesdropper, always sends a vacuum or
qubit to Bob. In another word, Bob's measurement is performed on a vacuum or qubit. Recently,
there are some works done in verifying the squash model
\cite{Koashi_NewModel_06,TT_Thres_08,BML_Squash_08}.

The third part of the QKD system is the source, which is the main concern of this paper. Here,
we consider the case that Eve has a full control of the photon source. That is, the source is
untrusted. This is a crucial assumption in the security proof of some QKD schemes, such as the
Plug$\&$Play system \cite{PaP_Gisin_02}. Recently, the QKD with untrusted source is studied
\cite{ZQL_untru_08,Wang_08}. By random sampling the photon source, the security is proved even
when Eve controls the photon source \cite{ZQL_untru_08}. This random sampling process requires
a fast random switch and a perfect ``intensity monitor". However, in reality, this is not
practical.

\begin{figure}[hbt]
\centering \resizebox{12cm}{!}{\includegraphics{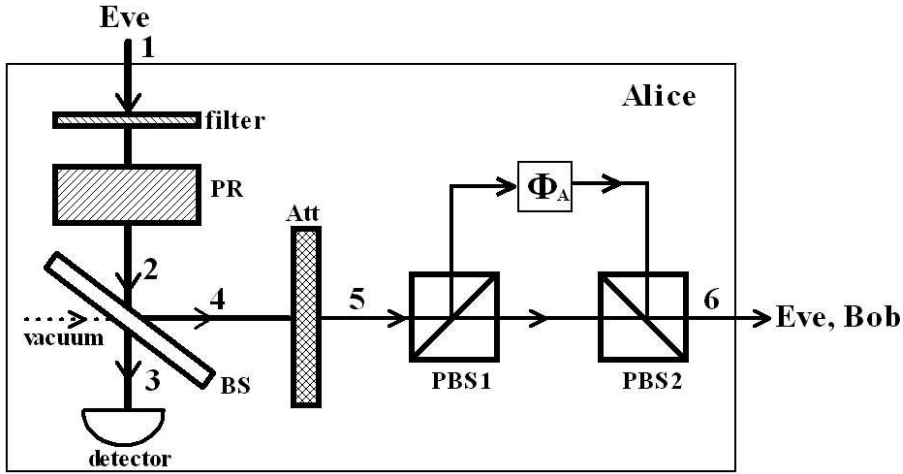}} \caption{A schematic diagram of the
setup on Alice's side. The untrusted photon source, prepared at P1 by Eve, passes through a
low-bandwidth filter and a phase randomizer (PR). Then a beam splitter (BS) (transmission:
$t_{bs}$) is used to separate it into two beams, 3 and 4. One beam goes to a photodetector
(detection efficiency: $t_{D}$) at P3 and the other is prepared for QKD at P4. An attenuator
(Att) between P4 and P5 has the attenuation coefficient $\eta_{s}$ ($\eta_{d}$) for the signal
(decoy) state. Two polarization beam splitters (PBS1 and PBS2) and a phase modulator
($\Phi_{A}$) between P5 and P6 are used for phase encoding.} \label{Fig:Setup}
\end{figure}

In this paper, we replace the random switch with a passive beam splitter. An inefficient
intensity monitor is used, which is modeled by a virtual beam splitter and an ideal detector.
The schematic diagram of the setup for Alice to monitor the photon source is shown in
Fig.~\ref{Fig:Setup}. In the following discussion, P$i$ with $i =1, \cdots, 6$, refers to
position $i$ in Fig.~\ref{Fig:Setup}.

The experimental procedure goes as follows.
\begin{enumerate} [(i)]
\item
First, the untrusted photon source passes through a filter, which guarantees the state in
single mode at P2. Then the global phase of the state is randomized by a phase randomizer (PR
in Fig.~\ref{Fig:Setup}). Thus, the state at P2 can be expressed as \cite{Decoy_05}
\begin{equation}
\rho_{2}= \sum_{N=0}^{\infty}P_t(N)|N\rangle\langle N|,
\end{equation}
where $P_t(N)$ satisfies $\sum_{N=0}^{\infty}P_t(N)=1$ and $|N\rangle$ is the number state.

\item
A small part ($1-t_{bs}$) of the beam at P2 will be reflected by the beam splitter (BS) to P4,
and the rest of the beam will be transmitted to an inefficient photodetector at P3. The
inefficient detector can be treated as another virtual beam splitter (transmission: $t_{D}$)
placed in front of an ideal detector. Thus, the photoelectron distribution $D(m)$ measured at
P3 is the Bernoulli transformation of the photon distribution $P_t(N)$ at P2. On the other
hand, $P_t(N)$ can be inferred by the inverse Bernoulli transformation of $D(m)$ \cite{Lee_93}
\begin{equation}\label{Untru:DmPN}
\begin{aligned}
D(m) &= B[P_t(N),\xi]=\sum_{N=m}^{\infty}P_t(N){N \choose m}\xi^{m}(1-\xi)^{N-m}, \\
P_t(N) &= B^{-1}[D(m),\xi^{-1}]=\sum_{m=N}^{\infty}D(m){m \choose N}\xi^{-N}(1-\xi^{-1})^{m-N},
\end{aligned}
\end{equation}
where $\xi=t_{bs}t_{D}$. When $\xi>0.5$, the $P_t(N)$ distribution can be efficiently recovered from $D(m)$ \cite{Kiss_95,Herzog_96}.

\item
From P2 to P4 and P4 to P5, the state will be changed by the BS and an attenuator (Att in
Fig.~\ref{Fig:Setup}) with an attenuation coefficient $1-t_{bs}$ and $\eta_{s}$ ($\eta_{d}$)
for signal state (decoy state), respectively. Thus, at P5, the photon number distribution
$P_{s}(i)$ for the signal state is $B[P_t(N),\eta{'_{s}}]$ (see Eq.~\eqref{Untru:DmPN}) and $P_{d}(i)=B[P_t(N),\eta{'_{d}}]$ for the decoy state, where $\eta{'_{s}}=\eta_{s}(1-t_{bs})$
and $\eta{'_{d}}=\eta_{d}(1-t_{bs})$.
\end{enumerate}

$D(m)$ can be measured directly from the experiment. Then $P_t(N)$ can be calculated by
Eq.~\eqref{Untru:DmPN}. From $P_t(N)$, we can bound $N\in[N_{min},N_{max}]$ with a confidence
($1-\varepsilon$). Following \cite{ZQL_untru_08}, one can lower bound the gain $Q_{1}^{s}$
($\underline{Q_{1}^{s}}$) and upper bound the error rate $e_{1}^{s}$ ($\overline{e_{1}^{s}}$)
of the single photon state. The secure key rate can be written as \cite{GLLP_04,ZQL_untru_08}
\begin{equation}\label{Untru:R}
R=q
\left[-Q_{s}f(E_{s})H_{2}(E_{s})+(1-\varepsilon)\underline{Q_{1}^{s}}(1-H_{2}(\overline{e_{1}^{s}}))\right],
\end{equation}
where $q$ is the basic reconciliation factor, $Q_{s}$ and $E_s$ are the overall gain and error
rate of the signal state, $f(E_{s})H_{2}(E_{s})$ is the leakage information in the error
correction (normally, $f(E_{s})\geq1$), and $H_{2}(x)=-x\log_{2}x-(1-x)\log_{2}(1-x)$ is the
binary entropy function.

\begin{figure}[hbt]
\centering \resizebox{12cm}{!}{\includegraphics{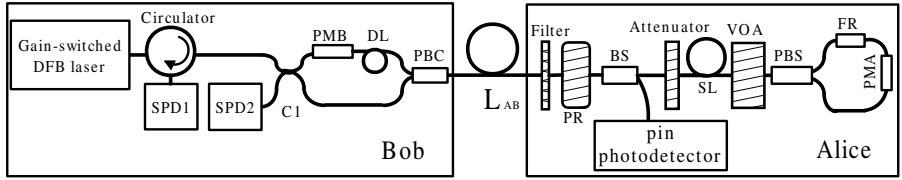}} \caption{Experimental setup for QKD
system. A gain-switched distributed feedback (DFB) laser diode emits a pulse train of laser
pulses [center wavelength: $1546.1~\rm nm$, pulse duration: $300~\rm ps$, pulse repetition
rate: 1~MHz] which are polarized by the polarization maintaining (PM) fiber's slow axis. The
circulator and 50/50 coupler (C1) are polarization maintained. SPD1, SPD2: single photon
detector; PBC/PBS: polarization beam combiner/splitter; PMA, PMB: phase modulator; DL: PM
fiber; $L_{AB}$: 25~km single-mode fiber; SL: 5~km storage line; VOA: variable optical
attenuator.  A 90$^{0}$ Faraday rotator (FR), together with the PBS, plays the same role as a
Faraday mirror. To implement the aforementioned scheme shown in Fig.~\ref{Fig:Setup}, we use a
filter (transparency window: $0.8~\rm{nm}$), phase randomizer (PR), 95/5 beam splitter (BS) and
pin photodetector (detection efficiency $t_{D}$: 0.8).} \label{Fig:Experiment}
\end{figure}

As shown in Fig.~\ref{Fig:Experiment}, we implement the aforementioned scheme in a standard
Plug$\&$Play QKD system. For applying the weak+vacuum decoy protocol, the experimental
parameters are listed in Table \ref{tab:para}. The average photon numbers for the signal state
and weak decoy state are $\mu$ and $\nu$, respectively. $N_{\mu}$, $N_{\nu}$ and $N_{0}$ are
the pulse numbers for the signal, weak decoy and vacuum states, respectively. Note that 50
laser pulses are set as a pulse train whose period is 350 ${\rm \mu s}$. In
Eq.~\eqref{Untru:R}, one can use $q=0.5F\cdot N_{\mu}/(N_{\mu}+N_{\nu}+N_{0})$, where $F=50{\rm
pulses}/350{\rm \mu s}$ is determined by the burst mode of laser source.  The temperature of
laser diode is well-controlled for decreasing intensity fluctuations.

\begin{table}[hbt]
  \centering
\begin{center}
\begin{tabular}{|c|c|c|c|c|c|c|c|c|}
\hline $\mu$ & $\nu$&$\xi$ &$N_{\mu}$&$N_{\nu}$&$N_{0}$ & $\eta{'_{s}}$ & $\eta{'_{d}}$&$\eta_{B}$ \\
\hline $0.48$& $0.06$&0.76&61747531&23056601&5712393 &$2.5\times10^{-8}$ & $3.1\times10^{-9}$&0.04\\
\hline
\end{tabular}
\caption{The experimental parameters. $\eta_{B}$ is the efficiency of Bob's detection system.
}\label{tab:para}
\end{center}
\end{table}

\begin{figure}[hbt]
\centering \resizebox{12cm}{!}{\includegraphics{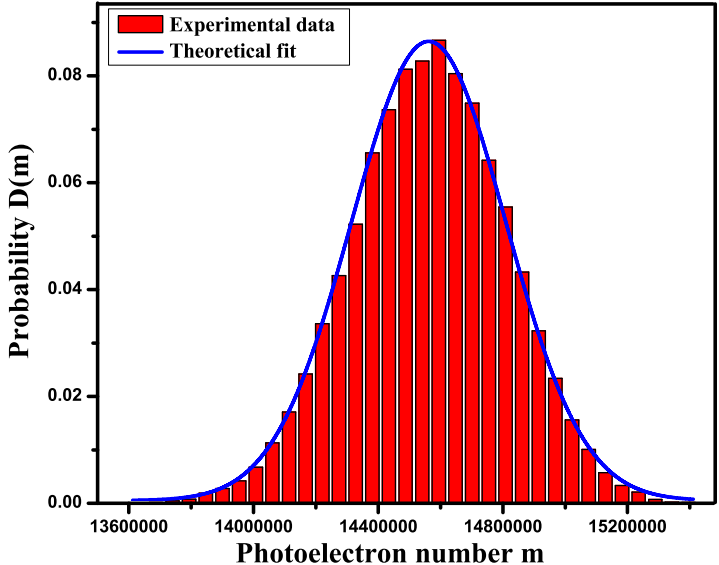}} \caption{(Color online) The
photoelectron probability distribution $D(m)$ with the experimental data shown as bars.  The
full curve represents the theoretical fit given that $P_{t}(N)$ is Gaussian distribution.
}\label{Fig:Result}
\end{figure}

To implement the scheme shown in Fig.~\ref{Fig:Setup}, we use a beam splitter (BS) with
$t_{bs}=0.95$ transmitting $95\%$ of the laser beam to a pin photodiode. In total,
$\xi=t_{bs}t_{D}=0.76$ of the incoming photon source is monitored by the photodiode. In the
detection, an integrating capacitor is charged by the emitted photoelectrons and its voltage is
proportional to the number of photoelectrons. The voltage signal is amplified, sent to a
sampling oscilloscope (Tektronix MSO4104) and recorded. Before next photoelectron pulse
arrives, due to the charge leakage, the capacitor is discharged.  $D(m)$ is experimentally
measured and shown in Fig.~\ref{Fig:Result}.

The experimental data are shown in Table \ref{tab:result}. $\langle m\rangle$ and $\langle
\Delta m^{2}\rangle$ are the average number and variance of photoelectrons $m$, respectively.
From Eq.~\eqref{Untru:DmPN}, one can derive that $\langle N\rangle=1.914\times10^{7}$ and
$\langle \Delta N^{2}\rangle=1.063\times10^{11}$ for $P_t(N)$, in fact, $\langle m\rangle=\xi
\langle N\rangle$ and $\langle \Delta m^{2}\rangle=\xi(1-\xi)\langle N\rangle+\xi^{2}\langle
\Delta N^{2}\rangle$. Fig.~\ref{Fig:Result} shows that $D(m)$ can be well fit as it is derived
from the Bernoulli transformation of a Gaussian distribution of $P_{t}(N)$.

\begin{table}[hbt]
  \centering
\begin{center}
\begin{tabular}{|c|c|c|c|c|c|c|}
\hline $Q_{s}$ & $Q_{d}$ & $Q_{0}$ & $E_{s}$&$E_{0}$&$\langle m\rangle$&$\langle\Delta m^{2}\rangle$ \\
\hline $5.84\times10^{-3}$& $7.48\times10^{-4}$ &$9.38\times10^{-5}$ & $2.1\%$&$46.1\%$&$1.455\times10^{7}$&$6.14\times10^{10}$\\
\hline
\end{tabular}
\caption{The experimental results.}\label{tab:result}
\end{center}
\end{table}

In the postprocessing, an improved ``Cascade" protocol \cite{Yamazaki_00} is applied for the
error correction and $f(E_{s})$ is estimated as 1.06 in Eq.~\eqref{Untru:R}. Based on the
Gaussian distribution of $P_t(N)$,  $N\in[N_{min}, N_{max}]$ with the confidence
($1-\varepsilon$) are chosen as $[1.751\times10^{7}, 2.077\times10^{7}]$ and
($1-5.7\times10^{-7}$), respectively. Then follow the analysis proposed in
Ref.~\cite{ZQL_untru_08}, $\underline{Q_{1}^{s}}$ and $\overline{e_{1}^{s}}$ are
$2.58\times10^{-3}$ and $3.77\%$, and the secure key rate can be calculated as $R\geq52$~bit/s
by Eq.~\eqref{Untru:R}. Note that with the same setup, the secure key rate is estimated as
$R\geq78$~bit/s if the source is trusted.

We have following remarks:
\begin{enumerate}
\item
Due to the computational complexity of the inverse Bernoulli transformation, see
Eq.~\eqref{Untru:DmPN}, we simply assume that $P_t(N)$ follows the Gaussian distribution. As
shown in Fig.~\ref{Fig:Result}, we deduce $D(m)$ from $P_t(N)$ and fit with experimental data.
Note that how to calculate Eq.~\eqref{Untru:DmPN} in postprocessing efficiently is an
interesting future topic.

\item
In the experiment, the electronic noise can be deducted from the signal to enhance the
estimation of $D(m)$. We use an oscilloscope to acquire data from the pin photodetector. The
data transmission speed of the oscilloscope limits the speed of photon source monitoring. In
the future, a high speed analog-to-digital circuit can be designed to replace the oscilloscope.

\item
In the analysis, we assume that the photon source is single-mode after passing through an ideal
filter. In the real experiment, the bandwidth of filter is not perfect. Thus, it is interesting
to investigate how to analyze a multi-mode photon source for QKD in future.

\item
Note that the statistical fluctuations from the finite data size \cite{Practical_05,Wang2_05}
and the accuracy of the estimation of $D(m)$ are not considered in our security analysis, which
are encouraged to be investigated in future.
\end{enumerate}

In conclusion, a practical method to monitor the photon statistics of QKD source is proposed
and is implemented in a real-life Plug\&Play QKD system. We run the experiment around 20
minutes. The final secure key rate is 52 bit/s, comparing to 78 bit/s when the source is
treated as a trusted source.

The fruitful discussions with H.-K.~Lo, N.~L\"{u}tkenhaus, B.~Qi and Y.~Zhao are greatly
appreciated. We gratefully acknowledge T. Liu for his work on the efficiency of the error
correction. This work is supported by the National Natural Science Foundation of China (Grant
No. 10474004) and the National Hi-Tech Program. X.~Ma acknowledges financial support from the
University of Toronto and the NSERC Innovation Platform Quantum Works.





\end{document}